\documentclass[
aps,nofootinbib,showpacs,showkeys,preprint
tightenlines,preprintnumbers,superscriptaddress]{revtex4}

\usepackage{epsf,epsfig,subfigure,graphicx,amsmath,amssymb}
\usepackage{color}
\newcommand{\dis}[1]{\begin{equation}\begin{split}#1\end{split}\end{equation}}

\def\lsim{\lower.7ex\hbox{$\;\stackrel{\textstyle<}{\sim}\;$}}

\newcommand{\gev}{\,\textrm{GeV}}

\newcommand{\etal}{{\it et al.}}
\newcommand{\Z}{{\bf Z}}

\newcommand{\ie}{{\it i.e.~}}

\newcommand{\NDW}{{N_{\rm DW}}}

\def\Qem{{$Q_{\rm em}$}}
\def\Qanom{{$Q_{\rm anom}$}}

\def\Z{{\bf Z}}

\def\EE{E$_8\times$E$_8^\prime$}

\def\SUflip{SU(5)$_{\rm flip}$}
\def\SUp{SU(5)$'$}
\def\Uanom{U(1)$_{\rm anom}$}

\def\Eo{E$_8$}
\def\Eh{E$_8'$}

\def\SMlike{{SU(3)$_c\times$SU(2)$_W\times$U(1)$^n$}}
\def\DSU5{{\it double SU(5)}}

\def\sixteen{\bf 16}

\def\two{{\bf 2}}
\def\fiveb{\overline{\bf 5}}
\def\five{{\bf 5}}
\def\tenb{\overline{\bf 10}}
\def\ten{{\bf 10}}
\def\one{{\bf 1}}

\begin{document}

\title{Calculation of axion--photon--photon coupling in string theory}
\author{Jihn E.  Kim }
\affiliation
{Department of Physics, Kyung Hee University, 26 Gyungheedaero, Dongdaemun-Gu, Seoul 130-701, Republic of Korea}

\begin{abstract}
The axion search experiments invite a plausible estimation of the axion--photon--photon coupling constant $\overline{c}_{a\gamma\gamma}$ in string models with phenomenologically acceptable visible sectors. We present the calculation of $\overline{c}_{a\gamma\gamma}$ with an exact Peccei-Quinn symmetry. In the Huh-Kim-Kyae $\Z_{12-I}$ orbifold compactification, we obtain
$\overline{c}_{a\gamma\gamma} =\frac{8}{3}$ even with a lot of singlets at the GUT scale, and the low-temperature axion search experiments will probe the QCD corrected coupling, 
$ {c}_{a\gamma\gamma} \simeq \overline{c}_{a\gamma\gamma}-1.98\simeq 0.69$.

\keywords{QCD axion, Axion-photon-photon coupling, $\Z_{12-I}$ orbifold compactification, Flipped SU(5) GUT}
\end{abstract}
\pacs{11.25.Mj, 14.80.Va, 98.80.Cq, 11.25.Wx}

\maketitle

\section{Introduction}
It seems that the Universe once passed the grand unification (GUT) scale energy region with its imprint survived until now \cite{BICEP2I}. If this BICEP2 result on the B-mode polarization survives on the matter of GUT scale energy density during inflation, it has a far-reaching implication in axion cosmology \cite{BCMaxion,KimRevs}.  Firstly, the implied high scale inflation nullifies the dilution idea of topological defects, strings and domain walls  of axion models \cite{KimDWsol14}. Second, if the QCD axion accounts for most of cold dark matter (CDM) in the Universe, the constraint from isocurvature perturbation rules out the anthropic region \cite{Pi84} of the axion parameter space \cite{Gondolo14}. If axion accounts for some fraction of CDM, then it may be possible to detect it by low temperature Sikivie-type detectors \cite{Sikivie}.
If we accept this high scale inflation scenario, there are two urgent issues to be clarified.

The first is to introduce the trans-Planckian value of inflaton, the so-called Lyth bound \cite{Lyth97}, within a well-motivated theory. Recently, Lyth argued for a rationale of any specific term working for a large $e$-fold number \cite{Lyth14}. There are three widely different classes of theories on this, the natural inflation completed with two nonabelian forces \cite{Freese90,KNP04}, appropriate quantum numbers under string-allowed discrete symmetries \cite{KimTrans14}, and M-flation \cite{Mflation}. Discrete symmetries are favored compared to global symmetries in string compactification \cite{KimPRL13}, which is thus welcome in obtaining a large $e$-folding by this method. If we rely on a single field inflation, it is generally very difficult to put the $e$-fold number at the bull's eye on the BICEP2 point \cite{KimTrans14}. However, there are some attempts to obtain the large $e$-folding from single field inflation \cite{ChoiKyae14}.

The second issue which motivated this paper is the domain wall problem in some axion models. Accepting high scale inflation, a string theory solution of the domain wall problem is possible \cite{KimDWsol14} using a discrete subgroup of the anomalous U(1) symmetry in string models. In string models with anomalous U(1) \cite{Witten84}, the model-independent (MI) axion becomes the longitudinal degree of the anomalous U(1) gauge boson, rendering it massive above $10^{16\,}\gev$ \cite{ChoiKim85}. Below $10^{16\,}\gev$, there results a global symmetry whose quantum numbers have descended from the original anomalous U(1) symmetry \cite{Kim88,Munoz89}. Thus, string models with the anomalous U(1) is suitable for introducing a spontaneously broken Peccei-Quinn (PQ) symmetry at the intermediate scale, to have an invisible axion \cite{KSVZ,DFSZ}. Now, because of the high scale inflation, it is a dictum to have the axion domain wall number one: $\NDW=1$. In string compactification, we found a solution of the domain wall problem \cite{KimDWsol14} by identifying vacua in terms of discrete subgroups of the anomalous U(1), which is the Choi-Kim (CK) mechanism \cite{ChoiKimDW85}.

The early (and so-far the only) example of the CK method using the anomalous U(1) was Ref. \cite{Kim88}, which however was based on a toy model. Here, we present the second example based on a phenomenologically acceptable grand unification (GUT) model from the heterotic string theory, leading to a $\NDW=1$ solution. In addition, we calculate the axion--photon-photon coupling strength, which is needed as a guideline in the axion detection experiments. It is in the Huh-Kim-Kyae (HKK) double SU(5) model \cite{HuhKK09,HKKprep09} from $\Z_{12-I}$ orbifold compactification. We may consider the $\Z_{12-I}$ compactification as the simplest one among the thirteen different orbifolds of the heterotic string \cite{ChoiKimBook}. One may be tempted to regard the $\Z_3$ orbifold compactification as the simplest one, but it is not so because the $\Z_3$ orbifold has twenty-seven fixed points while the $\Z_{12-I}$ orbifold has only three fixed points. If one follows the orbifold selection rules carefully, the $\Z_{12-I}$ orbifold compactification leads to the easiest way of obtaining a string model \cite{ChoiKimBook,KimKyaeZ12}. The most complicated orbifolds are from $\Z_{6-II}$ \cite{Raby04}. The \DSU5\, model is defined here as the model having three ($\tenb$ plus $\five$) families under one SU(5) and one ($\tenb'$ plus $\five'$) family under the other SU(5)$'$ toward a successful low energy supersymmetry (SUSY). One family SU(5)$'$ is needed for dynamical breaking of SUSY with confining force SU(5)$'$ \cite{DBSSU5}.\footnote{If one assumes gravity effects with gaugino condensation in SUSY breaking, one family SU(5)$'$ may not be needed \cite{Nilles}.}
There does not exist any \DSU5\, model in the $\Z_3$ orbifold compactification \cite{ChoiKimBook}, and we have not found any other \DSU5\, model yet beyond the HKK model in the computer scan of $\Z_{12-I}$ orbifolds.

Phenomenologically interesting orbifold models, in particular the standard-like models with gauge group \SMlike\, are interesting \cite{IKNQ87}, but for the study of anomalous U(1) they are too complicated because there are thirteen U(1) directions to consider. A simpler model with the GUT-type gauge coupling unification is the flipped-SU(5) GUT, \SUflip\,\cite{flipSU5}, in which a 16-dimensional set is obtained from the spinor representation {\bf 16} of SO(10). In this paper, the rank 5 gauge group SU(5)$\times$U(1)$_X$ is denoted as \SUflip. The fermionic construction of \SUflip\, was given in \cite{NAHE89}. The \DSU5~model contains \SUflip\,as the visible sector, and a successful phenomenology of the HKK model was discussed in Ref. \cite{HuhKK09}.

In Sec. \ref{Sec:Model}, we obtain the anomalous charge operator \Qanom~which is used for the PQ charges and list the charges for the \SUflip~nonsinglet representations. For the representations of the \Eh~sector nonabelian groups, the charges are listed in Appendix A. In Sec. \ref{Sec:axphcoup}, we list the charges for electromagnetically charged singlet representations and compute the axion--photon--photon coupling $\overline{c}_{a\gamma\gamma}$. Sec. \ref{Sec:Conclusion} is a conclusion.

\section{SU(5)$\times$U(1)$_X\times$SU(5)$'\times$\Uanom~ without domain wall problem}
\label{Sec:Model}

Recently, we emphasized that the early history of the Universe does not take
the possibility of inflating away the topological defects of axion models \cite{KimDWsol14}.
This implies that the axion solution of the strong CP problem via the spontaneous breaking of the Peccei-Quinn (PQ) symmetry is cosmologically disfavored if the axion domain wall number is not one \cite{SikivieDW82}. The solution by introducing $\NDW=1$ via the model-independent (MI) axion by the CK mechanism in string models is the following \cite{KimDWsol14}.
The MI axion has the anomaly coupling to gauge fields,
\dis{
\frac{a_{MI}}{32\pi^2 F_{MI}}\left( G\tilde G+ F_h\tilde F_h \right)
}
where $G\tilde G$ and $F_h\tilde F_h$ are the QCD and hidden sector anomalies, respectively. With the anomalous U(1)$_{\rm ga}$ gauge symmetry, below the U(1)$_{\rm ga}$ gauge boson scale a global symmetry survives and its spontaneous symmetry breaking allows the second axion coupling as
\dis{
\frac{ {\cal N}a_{2}}{32\pi^2 f_{2}}  G\tilde G
+\frac{ {\cal N}a_{2}}{32\pi^2 f_{2}} F_h\tilde F_h,
}
where ${\cal N}$ is common to  $G\tilde G$ and $F_h\tilde F_h$. Here, we assumed only one extra axion $a_2$ beyond the discrete subgroup of the MI axion direction. The fact that ${\cal N}$ is common to  $G\tilde G$ and $F_h\tilde F_h$ is essential to have a $\NDW=1$ solution. In this section, we show that indeed this is the case even though  $G\tilde G$ occurs from \Eo~and $F_h\tilde F_h$ occurs from \Eh. Identifying the same ${\cal N}$ is  the $\NDW=1$ solution via a discrete subgroup of \Uanom\,\cite{KimDWsol14}.

\begin{table}[!t]
{\tiny
\begin{center}
\begin{tabular}{|c|c|c |c||ccccccc|c|c|}
\hline Sect. & Colored states & SU(5)$_X$ &
Multiplicity & $Q_1$& $Q_2$ & $Q_3$ & $Q_4$ & $Q_5$ & $Q_6$ &  $Q_{\rm anom}$ & Label &$Q_a^{\gamma\gamma}$ \\[0.3em]
\hline\hline

$U$ & $\left(\underline{+\,+\,+\,-\,-\,};-\,-\,+\,
\right)(0^8)'$ & $\tenb_{-1}$     &  &--6 &--6 &+6 &0 &0 &0 &$-1638$ & $3C_2$&$-3276$ \\[0.3em]
$U$ & $\left(\underline{+\,-\,-\,-\,-\,};+\,-\,-\,
\right)(0^8)'$ & $\five_3$  &    &+6 &--6 &--6 &0 &0 &0 &$-126$ & $C_1$&$-294$ \\[0.3em]
\hline

$T_{4}^0$ &
$\left(\underline{+\,-\,-\,-\,-\,};\frac{-1}{6}\,\frac{-1}{6}\,\frac{-1}{6}\,
\right)(0^8)'$ &  $\five_3$   & $2$ &$-2$ &$-2$ &$-2$ &0 &0 &0 &$-378$& $2C_3$&$-882$ \\[0.5em]
$T_{4}^0$ &
$\left(\underline{+\,+\,+\,-\,-\,};\frac{-1}{6}\,\frac{-1}{6}\,\frac{-1}{6}\,
\right)(0^8)'$ &   $\tenb_{-1}$ & $2$ &$-2$ &$-2$ &$-2$ &0 &0 &0 &$-378,$ & $6C_4$ &$-756$\\[0.5em]
\hline

$T_{4}^0$ &
$\left(\underline{1\,0\,0\,0\,0\,};\frac{1}{3}\,\frac{1}{3}\,\frac{1}{3}\,
\right)(0^8)'$ &  $\five_{-2}$   & $2$ &$+4$ &$+4$ &$+4$ &0 &0 &0 &$+756$ & $2C_5$&$+1008$ \\[0.5em]
$T_{4}^0$ &
$\left(\underline{-1\,0\,0\,0\,0\,};\frac{1}{3}\,\frac{1}{3}\,\frac{1}{3}\,
\right)(0^8)'$ &   $\fiveb_{2}$ & $2$ &$+4$ &$+4$ &$+4$ &0 &0 &0 &$+756$ & $2C_6$ &$+1008$ \\[0.5em]
\hline

$T_{6}^0$ &
$\left(\underline{1\,0\,0\,0\,0\,};0\,0\,0\,
\right)(0^5; \frac{-1}{2}\,\frac{+1}{2}\,0)'$ &  $\five_{-2}$   & $3$ &0 &0 &0 &$-12$ &0 &0 &0
 & $3C_7$&0 \\[0.5em]
$T_{6}^0$ &
$\left(\underline{-1\,0\,0\,0\,0\,};0\,0\,0\,
\right)(0^5; \frac{+1}{2}\,\frac{-1}{2}\,0)'$ &   $\fiveb_{+2}$  & $3$ &0 &0 &0 & $+12$ &0 &0 &
0 & $3C_8$&0 \\[0.5em]
\hline

$T_{3}^0$ &$\left(\underline{+\,+\,+\,-\,-\,};0\,0\,0\,
\right)(0^5; \frac{-1}{4}\,\frac{-1}{4}\,\frac{+2}{4})'$ &  $\tenb_{-1}$ & $1$ &0 &0 &0 &0 &$+9$ &$+3$ &$-594,$& $3C_9$&$-1188$  \\[0.5em]
$T_{9}^0$ &
$\left(\underline{+\,+\,-\,-\,-\,};0\,0\,0\,
\right)(0^5; \frac{+1}{4}\,\frac{+1}{4}\,\frac{-2}{4} )'$ &   $\ten_{+1}$ & $1$ &0 &0 &0 &0 &$-9$ &$-3$ &$+594$ & $3C_{10}$&$+1188$ \\[0.5em]
\hline

$T_{7}^0$ &
$\left(\underline{-1\,0\,0\,0\,0\,};\frac{-1}{6}\,\frac{-1}{6}\,\frac{-1}{6}\,
\right)(0^5; \frac{-1}{4}\,\frac{-1}{4}\,\frac{+2}{4})'$ &  $\fiveb_{-2}$ & $1$ &$-2$ &$-2$ &$-2$ &0&$+9$ & $+3$&$-972$ & $C_{11}$ &$-1296$ \\[0.5em]
$T_{7}^0$ &
$\left(\underline{+1\,0\,0\,0\,0\,};\frac{-1}{6}\,\frac{-1}{6}\,\frac{-1}{6}\,
\right)(0^5; \frac{-1}{4}\,\frac{-1}{4}\,\frac{+2}{4})'$ &   $\five_{+2}$ & $1$ &$-2$ &$-2$  &$-2$ &0 &$+9$ &$+3$ &$-972$& $C_{12}$&$-1296$ \\[0.5em]
\hline\hline
& &  &$\sum_i Q(q_i)n(q_i)=$ &$-16$ &$-28$  &$+8$ &0 &$+18$ &$+6$&$-6984$ & $\sum_i =$ &$ -5406$  \\[0.3em]
\hline

\end{tabular}
\end{center}
\caption{The \SUflip~ states. Here, + represents $+\frac12$ and -- represents $-\frac12$. In the Label column, 3 is multiplied for $\ten$ and $\tenb$ each of which houses three quark and antiquarks. The PQ symmetry, being chiral, counts quark and antiquark in the same way.  The right-handed states in $T_3$ and $T_5$ are converted to the left-handed ones of $T_9$ and $T_7$,
respectively. 
}\label{tb:colorfields} }
\end{table}

In the ${\bf Z}_{12-I}$ HKK orbifold model, we have SU(5)$\times$U(1)$_X\times$SU(5)$'\times$\Uanom, and the key field contents under SU(5)$\times$SU(5)$'$ are $3\times  \sixteen+\{\ten, \tenb\}+\{\ten',\fiveb'\}$. The set $\{\ten,\tenb\}$ is needed for spontaneous breaking of SU(5)$_{\rm flip}\times$U(1)$_X$ down to the standard model gauge group. The set $\{\ten',\fiveb'\}$ is useful for SUSY breaking. Three copies of $\sixteen$ constitute three families of \SUflip.

The shift vector $V$ of $\Z_{12-I}$ is compsed of sixteen fractional numbers which are integer multiples of $\frac{1}{12}$, satisfying the modular invariance conditions. With the twist vector of the six internal dimensions with three complex numbers, $\phi=(\frac{5}{12},\frac{4}{12},\frac{1}{12})$, the condition is $12(V^2-\phi^2)=$ even integer. The Wilson line $W$ should satisfy the modular invariance conditions, $12(V^2-\phi^2)=$ even integer, $12V\cdot W=$ even integer, and $12W^2=$ even integer. The HKK model is \cite{HuhKK09},
\dis{
V&=\left(0~ 0~ 0~ 0~ 0~ \frac{-1}{6}~ \frac{-1}{6}~ \frac{-1}{6}\right)\left(0~ 0~ 0~ 0~ 0~ \frac{1}{4}~\frac{1}{4}~\frac{-2}{4}\right)'
\\[0.3em]
W &=\left(\frac23~ \frac23~ \frac23~ \frac23~ \frac23~ 0~\frac{-2}{3}~ \frac{2}{3}\right)\left(\frac23~ \frac23~ \frac23~ \frac23~ 0~ \frac{-2}{3}~ 0~0\right)'.
}

\begin{widetext}
In this model, the SU(5) charge raising and lowering generators are
\dis{
{\rm SU(5)}:F_a\,(a=1,\cdots,20) = \left(\underline{1\,-1\,0\,0\,0\,};\,0\,0\,0\right)(0^8)'  ,\label{eq:SU5p}
}
where the underline means permutations of the entries above the line.
The SU(5)$'$ charge raising and lowering generators are
\dis{
{\rm SU(5)}':\left\{ \begin{array}{l}
\Lambda_\alpha\,(\alpha=1,\cdots,12) =(0^8) \left(\underline{1\,-1\,0\,0}\,0\,;0\,0\,0\right)' ,\\
\Lambda_\alpha\,(\alpha=13,\cdots,16) =(0^8) \left(\underline{+\,+\,+\,-}\,+\,;-\,-\,-\right)',\\
\Lambda_\alpha\,(\alpha=17,\cdots,20) =(0^8) \left(\underline{+\,-\,-\,-}\,-\,;+\,+\,+\right)' .  \end{array}\right.\label{eq:SU5p}
}
The SU(2)$'$ charge raising and lowering generators are
\dis{
{\rm SU(2)}':\left\{ \begin{array}{l}
T^+  =(0^8) \left(+\,+\,+\,+\,+\,+\,+\,+\right)' ,\\
T^-  =(0^8) \left(-\,-\,-\,-\,-\,-\,-\,-\right)' .
\end{array} \right. \label{eq:SU2}
}
\end{widetext}
The rank 5 gauge group SU(5)$\times$U(1)$_X$ is denoted as \SUflip, where the hypercharge $Y_5\in$ SU(5) and $X$ are denoted as
\dis{
Y_5= \left(\frac{-1}{3}~\frac{-1}{3}~\frac{-1}{3}~\frac{+1}{2}~\frac{+1}{2}\,;0~\,0~\,0
\right)(0^8)',\label{eq:Y5charge}
}
\dis{
X= \left({-2\,-2\,-2\,-2\,-2\,};0~\,0~\,0
\right)(0^8)',\label{eq:Xcharge}
}
with the convention presented in Ref. \cite{ChoiKimBook}.
To get \Uanom, consider the rank 16 gauge group \SUflip$\times$U(1)$_1\times$U(1)$_2 \times$U(1)$_3$ from \Eo~and \SUp$\times$ SU(2)$'\times$U(1)$_4'\times$U(1)$_5'\times$U(1)$_6'$ from \Eh. The six U(1) charges are given by
\dis{
Q_1 &= \left({0^5\,};12~\,0~\,0\right)(0^8)' ,~ \tilde Q_1=\frac{1}{12}\, Q_1,\\
Q_2 &= \left({0^5\,};0~\,12~\,0\right)(0^8)' ,~ \tilde Q_2=\frac{1}{12}\, Q_2,\\
Q_3 &= \left({0^5\,};0~\,0~\,12\right)(0^8)' ,~ \tilde Q_3=\frac{1}{12}\, Q_3,\\
Q_4 &= (0^8)\left({0^4\,,0\,};12~\,-12~\,0\right)' ,~ \tilde Q_4=\frac{1}{12 \sqrt{2}}\, Q_4,\\
Q_5 &= (0^8)\left({0^5\,};-6~\,-6~\,~12\right)',~ \tilde Q_5=\frac{1}{6\sqrt{6}}\, Q_5,\\
Q_6 &= (0^8)\left({-6\,-6\,-6\,-6\,~~18\,};0\,~0\,~6\,\right)' , \, \tilde Q_6= \frac{1}{6\sqrt{14}}\, Q_6,\label{eq:defQs}
}
where tilded charges are the properly normalized U(1) charges, and norms of these charges are
\dis{
Q_1^2=Q_2^2 =Q_3^2 =144,~Q_4^2 =288 ,~Q_5^2 =216,~Q_6^2 =504.  \label{eq:Norms}
}

In Table \ref{tb:colorfields}, we list fields containing the standard model quarks (and anti-qaurks), where the U(1) charges are also shown. The PQ symmetry, being chiral, counts quark and antiquark in the same way, and we took into account the factor 3 for $\ten$ and $\tenb$ in the Label column.
The five anomaly free ${\rm U}(1)$s are
\dis{
P_1=\frac{1}{12\sqrt{5}}({Q}_1+2\,{Q}_3)&,~P_2=\frac{1}{6\sqrt{22}}(-{Q}_1+{Q}_2+2\,{Q}_6),~
P_3=\frac{1}{72}({Q}_5-3\,{Q}_6),\\
 P_4=&\frac{1}{12\sqrt{2}}Q_4,~P_5= \frac{1}{12\sqrt{74}}(3\,{Q}_3 -4\,{Q}_6).\label{eq:anfree}
}
The sixth U(1), which is orthogonal to Eq. (\ref{eq:anfree}) and  carries the anomaly, is
\dis{
{Q}_{\rm anom}&=84\,{Q}_1+147\,{Q}_2-42\,{Q}_3-63\,{Q}_5 -9\,{Q}_6 .  \label{eq:anom}
 }
For the nonabelian gauge groups from \Eh, we present two tables in Appendix A, Table \ref{tb:SUhfields} and Table \ref{tb:SU2fields}.
Comparing Table \ref{tb:colorfields} and Tables \ref{tb:SUhfields}  and \ref{tb:SU2fields} of Appendix A, we note that the anomaly sum of these U(1) charges are the same for three nonabelian groups, SU(5), SU(5)$'$ and SU(2)$'$.  In particular, the anomaly charges are the same, --6984, and we obtain the $\NDW=1$ solution as commented above \cite{KimDWsol14}. 
To compare with the $c_{a\gamma\gamma}$ column, we multiply --6984 by the index $\frac12$ of the fundamental representation of non-Abelian groups, \ie --3492

\section{Axion--photon--photon coupling}\label{Sec:axphcoup}
\begin{table}[!t]
{\tiny
\begin{center}
\begin{tabular}{|c|c|c|c||ccccccc|c|c|}
\hline Sect. & Singlet states &U(1)$_X$  &
Multiplicity & $Q_1$& $Q_2$ & $Q_3$ & $Q_4$ & $Q_5$ & $Q_6$ &  $Q_{\rm anom}$ & Label&$Q_a^{\gamma\gamma}$  \\[0.3em]
\hline\hline

$U$ & $\left( +\,+\,+\,+\,+\,;-\,+\,-\,
\right)(0^8)'$ & $\one_{-5}$  &  &$-6$ &$+6$ &$-6$ &0 &0 &0 &$+630$ & $S_1$&$+630$   \\[0.5em]
\hline

$T_{4}^0$ &
$\left( +\,+\,+\,+\,+\,;\frac{-1}{6}\,
\frac{-1}{6}\,\frac{-1}{6}\,\right)
\left(0^8\right)'$ &  $\one_{-5}$ & $2$ &$-2$ &$-2$ &$-2$ &$0$ &$0$ &$0$ &$-378$ & $2S_{24}$&$-378$ \\[0.5em]
\hline

$T_{4}^+$ &
$\left(\frac{1}{6}\,\frac{1}{6}\,\frac{1}{6}\,\frac{1}{6}\,\frac{1}{6}\,;\frac{-1}{6}\,
\frac{1}{6}\,\frac{1}{2}\,\right)
\left(\frac{1}{6}\,\frac{1}{6}\,\frac{1}{6}\,\frac{1}{6}\,\frac{-1}{2}\,;\frac{-1}{6}\,
\frac{-1}{2}\,\frac{1}{2}\,
\right)'$ &  $\one_{-5/3}$ & $2$ &$-2$ &+2 &+6 &$+4$ &$+10$ &$-10$ &$-666$ & $2S_2$&$-74$ \\[0.5em]
  &$\left(\frac{1}{6}\,\frac{1}{6}\,\frac{1}{6}\,\frac{1}{6}\,\frac{1}{6}\,;\frac{-1}{6}\,
\frac{1}{6}\,\frac{1}{2}\,\right)
\left(\frac{1}{6}\,\frac{1}{6}\,\frac{1}{6}\,\frac{1}{6}\,\frac{-1}{2}\,;\frac{-1}{6}\,
\frac{1}{2}\,\frac{-1}{2}\,
\right)'$ & $\one_{-5/3}$ & 2 & $-2$& $+2$ &$+6$ & $-8$ & $-8$ & $-16$ &$+522$  & $2S_3$   & $+58$\\[0.5em]
\hline

$T_{4}^-$ &
$\left(\frac{-1}{6}\,\frac{-1}{6}\,\frac{-1}{6}\,\frac{-1}{6}\,\frac{-1}{6}\,;\frac{-1}{6}\,
\frac{-1}{2}\,\frac{1}{6}\,\right)
\left(\frac{-1}{6}\,\frac{-1}{6}\,\frac{-1}{6}\,\frac{-1}{6}\,\frac{1}{2}\,;\frac{1}{6}\,
\frac{+1}{2}\,\frac{-1}{2}\,
\right)'$ &  $\one_{5/3}$ & $2$ &--2 &--6 &+2 &$-4$ &$-10$ &$+10$ &$-594$& $2S_4$&$-66$ \\[0.2em]
  &
$\left(\frac{-1}{6}\,\frac{-1}{6}\,\frac{-1}{6}\,\frac{-1}{6}\,\frac{-1}{6}\,;\frac{-1}{6}\,
\frac{-1}{2}\,\frac{1}{6}\,\right)
\left(\frac{-1}{6}\,\frac{-1}{6}\,\frac{-1}{6}\,\frac{-1}{6}\,\frac{1}{2}\,;\frac{1}{6}\,
\frac{-1}{2}\,\frac{+1}{2}\,
\right)'$ &$\one_{5/3}$&2  &$-2$ &$-6$  &$+2$  & $+8$ & $+8$ &$+16$ &$-1782$  & $2S_5$ &$-198$  \\[0.5em]
\hline

$T_{2}^+$ &
$\left(\frac{1}{3}\,\frac{1}{3}\,\frac{1}{3}\,\frac{1}{3}\,\frac{1}{3}\,;\frac{-1}{3}\,
\frac{1}{3}\,0\,\right)
\left(\frac{-1}{6}\,\frac{-1}{6}\,\frac{-1}{6}\,\frac{-1}{6}\,\frac{1}{2}\,;\frac{-1}{3}\,
0\,\frac{1}{2}\,
\right)'$ &  $\one_{-10/3}$ & $1$ &$-4$ &$+4$  &$0$&$-4$ &$+8$ &$+16$  &$-396$& $S_6$&$-176$ \\[0.5em]
  &
$\left(\frac{-1}{6}\,\frac{-1}{6}\,\frac{-1}{6}\,\frac{-1}{6}\,\frac{-1}{6}\,;\frac{1}{6}\,
\frac{-1}{6}\,\frac{1}{2}\,\right)
\left(\frac{-1}{6}\,\frac{-1}{6}\,\frac{-1}{6}\,\frac{-1}{6}\,\frac{1}{2}\,;\frac{2}{3}\,0\,
\frac{-1}{2}\,
\right)'$ &$\one_{5/3}$ &1  &$+2$ & $-2$ &$+6$ &$+8$  &$-10$  & $+10$ & $+162$ & $S_7$&$+18$   \\[0.5em]
  &
$\left(\frac{-1}{6}\,\frac{-1}{6}\,\frac{-1}{6}\,\frac{-1}{6}\,\frac{-1}{6}\,;\frac{1}{6}\,
\frac{-1}{6}\,\frac{1}{2}\,\right)
\left(\frac{-1}{6}\,\frac{-1}{6}\,\frac{-1}{6}\,\frac{-1}{6}\,\frac{1}{2}\,;\frac{-1}{3}\,0\,
\frac{1}{2}\,
\right)'$ &$\one_{5/3}$ &1 &$+2$ &$-2$&$+6$ &$-4$ &$+8$&$+16$ &$-1026$ & $S_8$&$-114$   \\[0.5em]
\hline

$T_{2}^-$ &
$\left(\frac{-1}{3}\,\frac{-1}{3}\,\frac{-1}{3}\,\frac{-1}{3}\,\frac{-1}{3}\,;\frac{-1}{3}\,
0\,\frac{1}{3}\,\right)
\left(\frac{1}{6}\,\frac{1}{6}\,\frac{1}{6}\,\frac{1}{6}\,\frac{-1}{2}\,;\frac{1}{3}\,
 0\,\frac{-1}{2}\,
\right)'$ &  $\one_{10/3}$  & $1$ &$-4$ &$0$ &$+4$ &$+4$ &$-8$ &$-16$ &$+144$& $S_9$&$+64$ \\[0.5em]
  &
$\left(\frac{1}{6}\,\frac{1}{6}\,\frac{1}{6}\,\frac{1}{6}\,\frac{1}{6}\,;\frac{1}{6}\,
\frac{-1}{2}\,\frac{-1}{6}\,\right)
\left(\frac{1}{6}\,\frac{1}{6}\,\frac{1}{6}\,\frac{1}{6}\,\frac{-1}{2}\,;\frac{-2}{3}\,0\,
\frac{1}{2}\,
\right)'$ &$\one_{-5/3}$ &1&$+2$ &$-6$ &$-2$&$-8$ & $+10$ &$-10$ &$-1170$ & $S_{10}$&$-130$ \\[0.5em]
  &
$\left(\frac{1}{6}\,\frac{1}{6}\,\frac{1}{6}\,\frac{1}{6}\,\frac{1}{6}\,;\frac{1}{6}\,
\frac{-1}{2}\,\frac{-1}{6}\,\right)
\left(\frac{1}{6}\,\frac{1}{6}\,\frac{1}{6}\,\frac{1}{6}\,\frac{-1}{2}\,;\frac{1}{3}\,0\,
\frac{-1}{2}\,
\right)'$ &$\one_{-5/3}$ &1&$+2$ &$-6$  &$-2$  &$+4$ &$-8$  & $-16$&$+18$ & $S_{11}$&$+2$ \\[0.5em]
\hline

$T_{1}^+$ &
$\left(\frac{-1}{3}\,\frac{-1}{3}\,\frac{-1}{3}\,\frac{-1}{3}\,\frac{-1}{3}\,;\frac{-1}{6}\,
\frac{1}{6}\,\frac{1}{2}\,\right)
\left(\frac{1}{6}\,\frac{1}{6}\,\frac{1}{6}\,\frac{1}{6}\,  \frac{-1}{2}\,;\frac{1}{12}\,\frac{-1}{4}\,0\,
\right)'$ &  $\one_{10/3}$ & $1$ &$-2$ &$+2$ &$+6$ &$+4$&$+1$ &$-13$ &$-72$ & $S_{12}$&$-32 $ \\[0.5em]
  &
$\left(\frac{1}{6}\,\frac{1}{6}\,\frac{1}{6}\,\frac{1}{6}\,\frac{1}{6}\,;\frac{-2}{3}\,
\frac{2}{3}\,0\,\right)
\left(\frac{1}{6}\,\frac{1}{6}\,\frac{1}{6}\,\frac{1}{6}\,\frac{-1}{2}\,;\frac{1}{12}\,
\frac{-1}{4}\,0\,
\right)'$ &$\one_{-5/3}$ &1  &$-8$ &$+8$&0 &$+4$&$+1$& $-13$&$+558$ & $S_{13}$&$+62$   \\[0.5em]
  &
$\left(\frac{1}{6}\,\frac{1}{6}\,\frac{1}{6}\,\frac{1}{6}\,\frac{1}{6}\,;\frac{1}{3}\,
\frac{-1}{3}\,0\,\right)
\left(\frac{1}{6}\,\frac{1}{6}\,\frac{1}{6}\,\frac{1}{6}\,\frac{-1}{2}\,;\frac{1}{12}\,
\frac{-1}{4}\,0\,
\right)'$ &$\one_{-5/3}$ &2 &$+4$ &$-4$&0& $+4$&$+1$&$-13$&$-198$ & $2S_{14}$&$-22$   \\[0.5em]
\hline

$T_{1}^-$ &
$\left(\frac{1}{3}\,\frac{1}{3}\,\frac{1}{3}\,\frac{1}{3}\,\frac{1}{3}\,;\frac{-1}{6}\,
\frac{1}{2}\,\frac{1}{6}\,\right)
\left(\frac{-1}{6}\,\frac{-1}{6}\,\frac{-1}{6}\,\frac{-1}{6}\,\frac{1}{2}\,;
\frac{5}{12}\,\frac{-1}{4}\,0\,
\right)'$ &  $\one_{-10/3}$ & $1$ &$-2$ &$+6$&$+2$& $+8$&$-1$ &$+13$&$+576$ & $S_{15}$&$+256$ \\[0.5em]
  &
$\left(\frac{-1}{6}\,\frac{-1}{6}\,\frac{-1}{6}\,\frac{-1}{6}\,\frac{-1}{6}\,;\frac{-2}{3}\,
0\,\frac{-1}{3}\, \right)
\left(\frac{-1}{6}\,\frac{-1}{6}\,\frac{-1}{6}\,\frac{-1}{6}\,\frac{1}{2}\,;\frac{5}{12}\,
\frac{-1}{4}\,0\,
\right)'$ &$\one_{5/3}$ &1  &$-8$&0&$-4$&$+8$&$-1$&$+13$&$-558$ & $S_{16}$&$-62$   \\[0.5em]
  &
$\left(\frac{-1}{6}\,\frac{-1}{6}\,\frac{-1}{6}\,\frac{-1}{6}\,\frac{-1}{6}\,;\frac{1}{3}\,
0\,\frac{2}{3}\,\right)
\left(\frac{-1}{6}\,\frac{-1}{6}\,\frac{-1}{6}\,\frac{-1}{6}\,\frac{1}{2}\,;\frac{5}{12}\,
\frac{-1}{4}\,0\,
\right)'$ &$\one_{5/3}$ &1 &$+4$ &0& $+8$&$+8$&$-1$&$+13$&$-54$ & $S_{17}$&$-6$   \\[0.5em]
\hline

$T_{7}^+$ &
$\left(\frac{-1}{3}\,\frac{-1}{3}\,\frac{-1}{3}\,\frac{-1}{3}\,\frac{-1}{3}\,;\frac{-1}{6}\,
\frac{1}{6}\,\frac{-1}{2}\,\right)
\left(\frac{1}{6}\,\frac{1}{6}\,\frac{1}{6}\,\frac{1}{6}\,\frac{-1}{2}\,;\frac{-5}{12}\,
\frac{1}{4}\,0\,
\right)'$ &  $\one_{-10/3}$  & $1$ &$-2$ &$+2$ &$-6$ & $-8$&$+1$&$-13$& $+432$& $S_{18}$&$+192$ \\[0.5em]
  &
$\left(\frac{1}{6}\,\frac{1}{6}\,\frac{1}{6}\,\frac{1}{6}\,\frac{1}{6}\,;\frac{1}{3}\,
\frac{2}{3}\,0\,\right)
\left(\frac{1}{6}\,\frac{1}{6}\,\frac{1}{6}\,\frac{1}{6}\,\frac{-1}{2}\,;\frac{-5}{12}\,
\frac{1}{4}\,0\,
\right)'$ & $\one_{5/3}$&1 &$+4$ &$+8$ &0& $-8$&$+1$ &$-13$ &$+1566$ &$S_{19}$ &$+174$   \\[0.5em]
  &
$\left(\frac{1}{6}\,\frac{1}{6}\,\frac{1}{6}\,\frac{1}{6}\,\frac{1}{6}\,;\frac{-2}{3}\,
\frac{-1}{3}\,0\,\right)
\left(\frac{1}{6}\,\frac{1}{6}\,\frac{1}{6}\,\frac{1}{6}\,\frac{-1}{2}\,;\frac{-5}{12}\,
\frac{1}{4}\,0\,
\right)'$ & $\one_{5/3}$&1 &$-8$ &$-4$ &0&$-8$&$+1$ &$-13$ &$-1206$ &  $S_{20}$&$-134$ \\[0.5em]
\hline

$T_{7}^-$ &
$\left(\frac{1}{3}\,\frac{1}{3}\,\frac{1}{3}\,\frac{1}{3}\,\frac{1}{3}\,;\frac{-1}{6}\,
\frac{-1}{2}\,\frac{1}{6}\,\right)
\left(\frac{-1}{6}\,\frac{-1}{6}\,\frac{-1}{6}\,\frac{-1}{6}\,\frac{1}{2}\,;
\frac{-1}{12}\,\frac{1}{4}\,0\,
\right)'$ &  $\one_{10/3}$  & $1$ &$-2$&$-6$& $+2$& $-4$& $-1$& $+13$& $-1188$& $S_{21}$&$-528$ \\[0.5em]
  &
$\left(\frac{-1}{6}\,\frac{-1}{6}\,\frac{-1}{6}\,\frac{-1}{6}\,\frac{-1}{6}\,;\frac{-2}{3}\,0\,
\frac{2}{3}\,\right)
\left(\frac{-1}{6}\,\frac{-1}{6}\,\frac{-1}{6}\,\frac{-1}{6}\,\frac{1}{2}\,;
\frac{-1}{12}\,\frac{1}{4}\,0\,
\right)'$  &$\one_{-5/3}$ &1 & $-8$&0&$+8$&$-4$ &$-1$ &$+13$ &$-1062$ & $S_{22}$&$-118$ \\[0.5em]
  &
$\left(\frac{-1}{6}\,\frac{-1}{6}\,\frac{-1}{6}\,\frac{-1}{6}\,\frac{-1}{6}\,;\frac{1}{3}\,0\,
\frac{-1}{3}\,\right)
\left(\frac{-1}{6}\,\frac{-1}{6}\,\frac{-1}{6}\,\frac{-1}{6}\,\frac{1}{2}\,;
\frac{-1}{12}\,\frac{1}{4}\,0\,
\right)'$  &$\one_{-5/3}$ &2 & $+4$&0&$-4$ &$-4$ &$-1$& $+13$&$+450$ & $S_{23}$& $+100$ \\[0.5em]

\hline\hline
& &  &$\sum_i Q (\one_i)n(\one_i)=$ &$-16$ &$-28$  &$+8$ &0&$+18$ &$+42$ &$-7632$ &$\sum_i =$ &$-1162$   \\[0.3em]
\hline

\end{tabular}
\end{center}
\caption{Electromagnetically charged singlets.  }\label{tb:ChSinglets}
}
\end{table}

For singlet fields, non-vanishing charges arise for non-vanishing $X$ quantum number of Eq. (\ref{eq:Xcharge}). Complete lists of the spectrum is found in the preprint version \cite{HKKprep09} of Ref. \cite{HuhKK09}. Singlets with non-vanishing $X$ charges are listed in Table \ref{tb:ChSinglets}. For the non-singlets, we also list the electromagnetic charges  in the last columns of Tables \ref{tb:colorfields}, \ref{tb:SUhfields}, and \ref{tb:SU2fields}. The electromagnetic charge \Qem~belongs to \SUflip, not depending on SU(5)$'$ and SU(2)$'$. The \SUflip~ assignments $(Y_5)_X$ are
\dis{
\five_3=\left(\begin{array}{c}  u^c\\   \nu_e\\  e^-\end{array}
\right)=\left(\begin{array}{c}  (\frac{-1}{3})_3\\[0.3em] (\frac{+1}{2})_3\\[0.3em]   (\frac{+1}{2})_3
\end{array}\right),~~
\tenb_{-1}=\left(\begin{array}{ccc}  & u& \\ d^c& & N\\ & d&    \end{array}
\right)=\left(\begin{array}{ccc}  &(\frac{-1}{6})_{-1}& \\[0.3em] (\frac23)_{-1}& & (-1)_{-1}\\[0.3em] &(\frac{-1}{6})_{-1}&    \end{array}
\right),~\one_{-5}=(e^+),
}
and we have the electromagnetic charge operator as
\dis{
Q_{\rm em}=W_3 +\frac{1}{5}Y_5- \frac{1}{5}X,
}
where $W_3$ is the third component of the weak isospin and the electroweak hypercharge is $Y=Y_5-\frac15 X$. Thus, the electromagnetic charges of the \SUflip~ representations are
\dis{
&\tenb_{-1}=\left((\frac{1}{3})_{\alpha},(\frac{2}{3})_{\alpha},(\frac{-1}{3})_{\alpha},0  \right),\, \ten_{+1}=\left((\frac{-1}{3})_{\alpha},(\frac{-2}{3})_{\alpha},(\frac{1}{3})_{\alpha},0  \right),\,\five_{3}=\left( (\frac{-2}{3})_{\alpha},0,-1\right),\,\one_X=(-\frac15 X),\\[0.6em]
&\five_{-2}=\left( ( \frac{1}{3})_{\alpha},\,1,\,0\right) ,\,
\fiveb_{+2}=\left( ( \frac{-1}{3})_{\alpha},\,0,\,-1\right)   ,\,
\fiveb_{-2}=\left( ( \frac{7}{15})_{\alpha},\,\frac{4}{5},\,\frac{-1}{5}\right)   ,\,
\five_{+2}=\left( ( \frac{-7}{15})_{\alpha},\,\frac{1}{5},\,\frac{-4}{5}\right),
}
where $\alpha$ is the color index and $-X/5$ for an \SUflip~ singlet is the electromagnetic charge of the singlet. For the \SUflip~ non-singlet representations, the traces are
\dis{
{\rm Tr\,}Q_{\rm em}^2(\tenb_{-1})=&{\rm Tr\,}Q_{\rm em}^2(\ten_{+1})=2,\,  {\rm Tr\,}Q_{\rm em}^2(\five_{+3})=\frac73,\, {\rm Tr\,}Q_{\rm em}^2(\one_{-5})=1,\\
{\rm Tr\,}Q_{\rm em}^2(\five_{-2})&=  {\rm Tr\,}Q_{\rm em}^2(\fiveb_{+2})=
{\rm Tr\,}Q_{\rm em}^2(\fiveb_{-2})= {\rm Tr\,}Q_{\rm em}^2(\five_{+2})=\frac43.\label{eq:QemTr}
}
In passing, note that the trace of $Q_{\rm em}^2$ for an anomaly-free irreducible set, including the fundamental representation of GUT representations, defines $\sin^2\theta_W^0$ of that GUT. Such examples in  Eq. (\ref{eq:QemTr}) are $\tenb_{-1}+\five_{+3}+\one_{-5}, \five_{-2}+\fiveb_{+2},$ etc. Assuming the universal coupling for all gauge groups in string theory, from $ \five_{-2}+\fiveb_{+2}$ for example, we obtain
\dis{
\sin^2\theta_W^0=\frac{{\rm Tr\,}W_3^2}{{\rm Tr\,}Q_{\rm em}^2}=\frac38.
}

From the last column of Tables \ref{tb:colorfields}, \ref{tb:ChSinglets},  \ref{tb:SUhfields}, and \ref{tb:SU2fields}, we obtain 
${\rm Tr\,} Q_a^{\gamma\gamma} Q_{\rm em}^2 =- 9312$.  Thus, we obtain
\dis{
\overline{c}_{a\gamma\gamma}= 
\frac{-9312}{-3492} =\frac{8}{3},\label{eq:number}
}
which is the same as the $(d^c,e)$ unification model \cite{Kim98}.
With the chiral symmetry breaking effect, $-1.98$, calculated with $m_u/m_d\simeq 0.5$ \cite{ManoharPDG}, we obtain $ {c}_{a\gamma\gamma} \simeq \overline{c}_{a\gamma\gamma}-1.98\simeq 0.69$. 
The cavity detector probes the axion--photon--photon coupling in a strong magnetic field {\bf B},
\dis{
{\cal L}=c_{a\gamma\gamma}\frac{\alpha_{\rm em}\, a}{8\pi \,f_a}\,{\bf E}\cdot{\bf B}.
}

\section{Conclusion}\label{Sec:Conclusion}

We computed the axion--photon--photon coupling in a phenomenologically viable HKK \SUflip$\times$SU(5)$'\times $U(1)$_{\rm anom}$ model from the heterotic \EE~string compactified on the $\Z_{12-I}$ orbifold, $\overline{c}_{a\gamma\gamma} = {8}/{3}$  even with a lot of singlets at the GUT scale, leading to ${c}_{a\gamma\gamma}^2\simeq 0.47$. It is exactly the $(d^c,e)$ unification value \cite{Kim98} in the DFSZ model, $ {c}_{a\gamma\gamma}^2\simeq 0.47$.
There has appeared a $ {c}_{a\gamma\gamma}^2$ calculation with an approximate Peccei-Quinn symmetry before \cite{ChoiKimKim06}, resulting with a smaller $ {c}_{a\gamma\gamma}^2(\simeq 0.07)$ compared to the present value, and it is likely that $ {c}_{a\gamma\gamma}^2$ from string takes some range of parameters. The present calculation is the first calculation with an exact PQ symmetry with the anomalous U(1). If the anomalous U(1) descends down for the QCD axion, it is likely that  $\overline{c}_{a\gamma\gamma} $ is $\frac83$.

\section*{Appendix A}
In this Appendix, we list the charges of the \Eh~nonabelian group representations, those of SU(5)$'$ in Table \ref{tb:SUhfields} and those of SU(2)$'$ in Table \ref{tb:SU2fields}. As claimed, the anomalous charges are exactly the same as that of the visible sector group SU(5), --6984. These hidden sector particles can carry the electromagnetic charges and they contribute to the coupling $\overline{c}_{a\gamma\gamma}$.

\begin{table}[h]
{\tiny
\begin{center}
\begin{tabular}{|c|c|c|c||ccccccc|c|c|}
\hline Sect. & States & SU(5)$'$ & Multiplicity & $Q_1$& $Q_2$ & $Q_3$ & $Q_4$ & $Q_5$ & $Q_6$ &  $Q_{\rm anom}$ & Label &$Q_a^{\gamma\gamma}$  \\[0.3em]
\hline\hline

$T_{1}^0$ &
$\left(1\,0\,0\,0\,0\,;\frac{-1}{6}\,\frac{-1}{6}\,\frac{-1}{6}\,\right)
(\underline{-1\,0\,0\,0\,}\,0\,; \frac{1}{4}\,\frac{1}{4}\,\frac{1}{2} )'$ &  $\tenb'_0$  & $1$ &$-2$ &$-2$ &$-2$ &0 &$+3$ & $+9$ &$-648$ & $3T_1'$&0   \\[0.5em]
 & $\left(1\,0\,0\,0\,0\,;\frac{-1}{6}\,\frac{-1}{6}\,\frac{-1}{6}\,\right)
(\underline{\frac{1}{2}\,\frac{1}{2}\,\frac{-1}{2}\,\frac{-1}{2}\,}\,\frac{1}{2}\,; \frac{-1}{4}\,\frac{-1}{4}\,0 )'$ &      &&&&&&&& & & \\[0.5em]
\hline
$T_{1}^0$ &$\left(0\,0\,0\,0\,0\,;\frac{-1}{6}\,\frac{-1}{6}\,\frac{-1}{6}\,\right)
(\underline{1\,0\,0\,0\,}\,0\,; \frac{1}{4}\,\frac{1}{4}\,\frac{1}{2} )'$ &  $(\five',\two')_0$   & $1$ &$-2$ &--2 &--2 &0  &$+3$ &$-3$ &$-540$ & $2F_1'$&0 \\[0.2em]
&$\left(0\,0\,0\,0\,0\,;\frac{-1}{6}\,\frac{-1}{6}\,\frac{-1}{6}\,\right)
({0\,0\,0\,0\,}\,0\,; \frac{-3}{4}\,\frac{-3}{4}\,\frac{-1}{2} )'$ &  &&&&&&&& &&   \\[0.2em]
&$\left(0\,0\,0\,0\,0\,;\frac{-1}{6}\,\frac{-1}{6}\,\frac{-1}{6}\,\right)
(\underline{\frac12\,\frac{-1}{2}\,\frac{-1}{2}\,\frac{-1}{2}\,}\,\frac{-1}{2}\,\,; \frac{-1}{4}\,\frac{-1}{4}\,0 )'$ &  &&&&&&&& &&   \\[0.2em]
&$\left(0\,0\,0\,0\,0\,;\frac{-1}{6}\,\frac{-1}{6}\,\frac{-1}{6}\,\right)
({\frac12\,\frac{1}{2}\,\frac{1}{2}\,\frac{1}{2}\,}\,\frac{-1}{2}\,\,; \frac{-1}{4}\,\frac{-1}{4}\,0 )'$ &  &&&&&&&& &&   \\[0.6em]
\hline
$T_{1}^0$&$\left(0\,0\,0\,0\,0\,;\frac{-1}{6}\,\frac{-1}{6}\,\frac{-1}{6}\,\right)
(\underline{\frac{-1}{2}\,\frac{1}{2}\,\frac{1}{2}\,\frac{1}{2}\,}\,\frac{-1}{2}\,\,; \frac{-1}{4}\,\frac{-1}{4}\,0 )'$ & $\fiveb'_0$  & $1$ &--2 &--2 &--2 &0 &$+3$ &$-15$ &$-432$ &  $F_2'$ &0  \\[0.2em]
&$\left(0\,0\,0\,0\,0\,;\frac{-1}{6}\,\frac{-1}{6}\,\frac{-1}{6}\,\right)
(0\,0\,0\,0\,-1 \,; \frac{1}{4}\,\frac{1}{4}\,\frac{1}{2} )'$ && &&&&&&&&& \\[0.3em]
\hline

$T_{1}^+$
&$\left(\frac{1}{6}\,\frac{1}{6}\,\frac{1}{6}\,\frac{1}{6}\,\frac{1}{6}\,;\frac{1}{3}\,
\frac{-1}{3}\,0 \,\right)
\left(\underline{\frac{-5}{6}\,\frac{1}{6}\,\frac{1}{6}\,\frac{1}{6}\,}\,\frac{1}{2}\,
;\frac{1}{12}\,\frac{-1}{4}\,0\,\right)'$ &  $\fiveb'_{-5/3}$   & $1$ &$+4$ &$-4$ &0 &$+4$
&$+1$ &$+11$ &$-414$ & $F'_3$&$-230$ \\[0.2em]
&$\left(\frac{1}{6}\,\frac{1}{6}\,\frac{1}{6}\,\frac{1}{6}\,\frac{1}{6}\,;\frac{1}{3}\,
\frac{-1}{3}\,0 \,\right)
\left(\frac{-1}{3}\,\frac{-1}{3}\,\frac{-1}{3}\,\frac{-1}{3}\,0\,;\frac{7}{12}\,
\frac{1}{4}\,\frac{1}{2}\,\right)'$ &&&&&&&&& &&  \\[0.3em]
\hline

$T_{4}^+$ &
$\left(\frac{1}{6}\,\frac{1}{6}\,\frac{1}{6}\,\frac{1}{6}\,\frac{1}{6}\,;\frac{-1}{6}\,
\frac{1}{6}\,\frac{1}{2}\,\right)
\left(\underline{\frac{2}{3}\,\frac{-1}{3}\,\frac{-1}{3}\,\frac{-1}{3}\,}\,0\,;
\frac{1}{3}\,0\,0\,\right)'$ &  $\five'_{-5/3}$  & $3$ &--2 &+2 &+6 &$+4$ &$-2$ &$+2$ &$-18$ & $3F'_4$&$-10$ \\[0.2em]
 &$\left(\frac{1}{6}\,\frac{1}{6}\,\frac{1}{6}\,\frac{1}{6}\,\frac{1}{6}\,;
 \frac{-1}{6}\,\frac{1}{6}\,\frac{1}{2}\,\right)
\left(\frac{1}{6}\,\frac{1}{6}\,\frac{1}{6}\,\frac{1}{6}\,\frac{1}{2}\,;
\frac{-1}{6}\,\frac{-1}{2}\,\frac{-1}{2}\,\right)'$ & && &&&&&& & &   \\[0.2em]
\hline
$T_{4}^-$ &
$\left(\frac{-1}{6}\,\frac{-1}{6}\,\frac{-1}{6}\,\frac{-1}{6}\,\frac{-1}{6}\,;
\frac{-1}{6}\,\frac{-1}{2}\,\frac{1}{6}\,\right)
\left(\underline{\frac{-2}{3}\,\frac{1}{3}\,\frac{1}{3}\,\frac{1}{3}\,}\,0\,;
\frac{-1}{3}\,0\,0\,\right)'$ &  $\fiveb'_{5/3}$   & $3$ &--2 &--6 &+2 &$-4$
&$+2$ &$-2$ &$-1242$& $3F'_5$&$-690$ \\
 &$\left(\frac{-1}{6}\,\frac{-1}{6}\,\frac{-1}{6}\,\frac{-1}{6}\,\frac{-1}{6}\,;
 \frac{-1}{6}\,\frac{-1}{2}\,\frac{1}{6}\,\right)
\left(\frac{-1}{6}\,\frac{-1}{6}\,\frac{-1}{6}\,\frac{-1}{6}\,\frac{-1}{2}\,;
\frac{1}{6}\,\frac{1}{2}\,\frac{1}{2}\,\right)'$ & && &&&&&& & &  \\[0.2em]
\hline

$T_{7}^-$
&$\left(\frac{-1}{6}\,\frac{-1}{6}\,\frac{-1}{6}\,\frac{-1}{6}\,\frac{-1}{6}\,;
\frac{1}{3}\,0 \,\frac{-1}{3}\,\right)
\left(\underline{\frac{5}{6}\,\frac{-1}{6}\,\frac{-1}{6}\,\frac{-1}{6}\,}\,\frac{-1}{2}\,;
\frac{-1}{12}\,\frac{1}{4}\,0\,\right)'$ &  $\five'_{-5/3}$ & $1$ &+4 &0 &--4 &$-4$
&$-1$ &$-11$  &$+666$& $F'_6$&$+370$ \\[0.2em]
&$\left(\frac{-1}{6}\,\frac{-1}{6}\,\frac{-1}{6}\,\frac{-1}{6}\,\frac{-1}{6}\,;
\frac{1}{3}\,0 \,\frac{-1}{3}\,\right)
\left(\frac{1}{3}\,\frac{1}{3}\,\frac{1}{3}\,\frac{1}{3}\,0\,;\frac{-7}{12}\,
\frac{-1}{4}\,\frac{-1}{2}\,\right)'$ & &&&&&&&& &&  \\[0.3em]
\hline\hline
&   &&$\sum_i Q(q_i')n(q_i')=$ &$-16$ &$-28$  &$+8$ &0 &$+18$ &$+6$&$-6984$ & $\sum_i
=$&$ -1960$   \\[0.2em]
\hline

\end{tabular}
\end{center}
\caption{The SU(5)$'$ representations.  Notations are the same as in Table \ref{tb:colorfields}.  }\label{tb:SUhfields} }
\end{table}

\begin{table}[h]
{\tiny
\begin{center}
\begin{tabular}{|c|c|c|c||ccccccc|c|c|}
\hline Sect. & States & SU(2)$'$  & Multiplicity & $Q_1$& $Q_2$ & $Q_3$ & $Q_4$ & $Q_5$ & $Q_6$ &  $Q_{\rm anom}$ & Label&$Q_a^{\gamma\gamma}$  \\[0.3em]
\hline\hline

$T_1^0$
&$\left(0\,0\,0\,0\,0\,;\frac{-1}{6}\,\frac{-1}{6}\,\frac{-1}{6}\,\right)
(\underline{1\,0\,0\,0\,}\,0\,; \frac{1}{4}\,\frac{1}{4}\,\frac{1}{2} )'$ &  $(\five',\two')_0$   & $1$ &--2 &--2 &--2 &0 &$+3$ &$-3$ &$-540$& $5D_1'$ &Considered\\[0.2em]
&$\left(0\,0\,0\,0\,0\,;\frac{-1}{6}\,\frac{-1}{6}\,\frac{-1}{6}\,\right)
({0\,0\,0\,0\,}\,0\,; \frac{-3}{4}\,\frac{-3}{4}\,\frac{-1}{2} )'$ &&&&&&&&& &&in Table \ref{tb:SUhfields}   \\[0.5em]
\hline

$T_{1}^0$ &
$\left(0\,0\,0\,0\,0\,;\frac{-1}{6}\,
\frac{-1}{6}\,\frac{-1}{6}\,\right)
\left(0\,0\,0\,0\,1\,;\frac{1}{4}\,\frac{1}{4}\,\frac{1}{2}\,
\right)'$ &  $\two'_0$  & $1$ &$-2$ &$-2$ &$-2$ &0 &$+3$ &$+21$ &$-756$& $D_2$&0 \\[0.5em]
\hline

$T_{1}^+$ &
$\left(\frac{1}{6}\,\frac{1}{6}\,\frac{1}{6}\,\frac{1}{6}\,\frac{1}{6}\,;\frac{1}{3}\,
\frac{-1}{3}\,0\,\right)
\left(\frac{1}{6}\,\frac{1}{6}\,\frac{1}{6}\,\frac{1}{6}\,\frac{1}{2}\,;\frac{1}{12}\,
\frac{3}{4}\,0\,
\right)'$
&$\two'_{-5/3}$  & $1$ &$+4$ &$-4$ &$0$ &$-8$ &$-5$ &$+5$ &$+18$& $D_3$&$+4$ \\[0.5em]
\hline

$T_{1}^-$ &
$\left(\frac{-1}{6}\,\frac{-1}{6}\,\frac{-1}{6}\,\frac{-1}{6}\,\frac{-1}{6}\,;\frac{-2}{3}\,
0\,\frac{-1}{3}\,\right)
\left(\frac{1}{3}\,\frac{1}{3}\,\frac{1}{3}\,\frac{1}{3}\,0\,;\frac{-1}{12}\,
\frac{1}{4}\,\frac{1}{2}\,\right)'$
&$\two'_{5/3}$  & $1$ &$-8$ &$0$ &$-4$ &$-4$ &$+5$ &$-5$ &$-774$ & $D_4$& $-172$ \\[0.5em]
\hline

$T_{1}^-$ &
$\left(\frac{-1}{6}\,\frac{-1}{6}\,\frac{-1}{6}\,\frac{-1}{6}\,\frac{-1}{6}\,;\frac{1}{3}\,
0\,\frac{2}{3}\,\right)
\left(\frac{1}{3}\,\frac{1}{3}\,\frac{1}{3}\,\frac{1}{3}\,0\,;\frac{-1}{12}\,
\frac{1}{4}\,\frac{1}{2}\,\right)'$
&$\two'_{5/3}$  & $1$ &$+4$ &$0$ &$+8$ &$-4$ &$+5$ &$-5$ &$-270$ & $D_5$& $-60$ \\[0.5em]
\hline

$T_{2}^+$ &
$\left(\frac{-1}{6}\,\frac{-1}{6}\,\frac{-1}{6}\,\frac{-1}{6}\,\frac{-1}{6}\,;\frac{1}{6}\,
\frac{-1}{6}\,\frac{1}{2}\,\right)
\left(\frac{1}{3}\,\frac{1}{3}\,\frac{1}{3}\,\frac{1}{3}\,0\,;\frac{1}{6}\,\frac{1}{2}\,0\,
\right)'$ &  $\two'_{5/3}$ & $1$ &$+2$ &$-2$  &$+6$&$-4$ &$-4$ &$-8$  &$-54$& $D_6$& $-12$  \\[0.5em]
\hline

$T_{2}^-$ &
$\left(\frac{1}{6}\,\frac{1}{6}\,\frac{1}{6}\,\frac{1}{6}\,\frac{1}{6}\,;\frac{1}{6}\,
\frac{-1}{2}\,\frac{-1}{6}\,\right)
\left(\frac{1}{6}\,\frac{1}{6}\,\frac{1}{6}\,\frac{1}{6}\,\frac{1}{2}\,;\frac{1}{3}\,0\,
\frac{1}{2}\,
\right)'$ &  $\two'_{-5/3}$  & $1$ &$+2$ &$-6$ &$-2$ &$+4$ &$+4$ &$+8$ &$-954$& $D_7$& $-212$  \\[0.5em]
\hline

$T_{4}^+$ &
$\left(\frac{1}{6}\,\frac{1}{6}\,\frac{1}{6}\,\frac{1}{6}\,\frac{1}{6}\,;\frac{-1}{6}\,
\frac{1}{6}\,\frac{1}{2}\,\right)
\left(\frac{1}{6}\,\frac{1}{6}\,\frac{1}{6}\,\frac{1}{6}\,\frac{1}{2}\,;\frac{-1}{6}\,
\frac{1}{2}\,\frac{1}{2}\,
\right)'$ &  $\two'_{-5/3}$ & $2$ &--2 &+2 &+6 &$-8$ &$+4$ &$+8$ &$-450$ & $2D_8$& $-100$ \\[0.5em]
\hline

$T_{4}^-$ &
$\left(\frac{-1}{6}\,\frac{-1}{6}\,\frac{-1}{6}\,\frac{-1}{6}\,\frac{-1}{6}\,;\frac{-1}{6}\,
\frac{-1}{2}\,\frac{1}{6}\,\right)
\left(\frac{1}{3}\,\frac{1}{3}\,\frac{1}{3}\,\frac{1}{3}\,0\,;\frac{2}{3}\,0\,0\,
\right)'$ &  $\two'_{5/3}$ & $2$ &--2 &--6 &+2 &$+8$ &$-4$ &$-8$ &$-810$& $2D_9$& $-180$ \\[0.5em]
\hline

$T_{7}^+$ &
$\left(\frac{1}{6}\,\frac{1}{6}\,\frac{1}{6}\,\frac{1}{6}\,\frac{1}{6}\,;\frac{1}{3}\,
\frac{2}{3}\,0\,\right)
\left(\frac{1}{6}\,\frac{1}{6}\,\frac{1}{6}\,\frac{1}{6}\,\frac{1}{2}\,;\frac{7}{12}\,
\frac{1}{4}\,0\,
\right)'$ &  $\two'_{5/3}$  & $1$ &$+4$ &$+8$ &$0$ &$+4$ &$-5$ &$+5$ &$+1782$& $D_{10}$& $+396$ \\[0.5em]
\hline

$T_{7}^+$ &
$\left(\frac{1}{6}\,\frac{1}{6}\,\frac{1}{6}\,\frac{1}{6}\,\frac{1}{6}\,;\frac{-2}{3}\,
\frac{-1}{3}\,0\,\right)
\left(\frac{1}{6}\,\frac{1}{6}\,\frac{1}{6}\,\frac{1}{6}\,\frac{1}{2}\,;\frac{7}{12}\,
\frac{1}{4}\,0\,
\right)'$ &  $\two'_{5/3}$  & $1$ &$-8$ &$-4$ &$0$ &$+4$ &$-5$ &$+5$ &$-990$& $D_{11}$& $-220$ \\[0.5em]
\hline

$T_{7}^-$ &
$\left(\frac{-1}{6}\,\frac{-1}{6}\,\frac{-1}{6}\,\frac{-1}{6}\,\frac{-1}{6}\,;\frac{1}{3}\,0\,
\frac{-1}{3}\,\right)
\left(\frac{1}{3}\,\frac{1}{3}\,\frac{1}{3}\,\frac{1}{3}\,0\,;\frac{5}{12}\,\frac{-1}{4}\,
\frac{1}{2}\,
\right)'$ &  $\two'_{-5/3}$  & $1$ &$+4$&$0$ &$-4$ &$+8$ &$+5$ &$-5$ &$+234$& $D_{12}$& $+52$ \\[0.5em]
\hline\hline
& &  &$\sum_i Q(\two_i')n(\two_i')=$ &$-16$ &$-28$  &$+8$ &0&$+18$ &$+6$ &$-6984$ &$\sum_i =$ & $-784$  \\[0.3em]
\hline

\end{tabular}
\end{center}
\caption{The SU(2)$'$ representations. Notations are the same as in Table \ref{tb:colorfields}. We listed only the upper component of SU(2)$'$ from which the lower component can be obtained by applying $T^-$ of Eq. (\ref{eq:SU2}).  }\label{tb:SU2fields}
}
\end{table}


\vskip 0.2cm

\noindent {\bf Acknowledgments}: {I thank Johar Ashfaque and Bumseok Kyae for pointing out errors in the table entries of v1.
This work is supported in part by the National Research Foundation (NRF) grant funded by the Korean Government (MEST) (No. 2005-0093841) and by the IBS(IBS CA1310).
}

\end{document}